\documentclass[aps,pra,twocolumn,superscriptaddress,floatfix]{revtex4}

\usepackage{graphicx}
\usepackage{amssymb}
\usepackage{subfigure}
\usepackage{amsmath}
\usepackage{amsfonts}
\usepackage{natbib}
\usepackage{bm}
\usepackage{color}
\usepackage{placeins}

\begin{document}

\title{Remote preparation of single photon vortex thermal states}

\author{T. H\"{a}ffner}
\affiliation{Departamento de F\'{i}sica, Universidade Federal de Santa Catarina, CEP 88040-900, Florian\'{o}plis, SC, Brazil}
\author{G. L. Zanin}
\affiliation{Departamento de F\'{i}sica, Universidade Federal de Santa Catarina, CEP 88040-900, Florian\'{o}plis, SC, Brazil}
\author{R. M. Gomes}
\affiliation{Institute of Physics, Federal University of Goi\'{a}s, 74690-900, Goi\^{a}nia, GO, Brazil}
\author{L. C. C\'{e}leri}
\email{lucas@chibebe.org}
\affiliation{Institute of Physics, Federal University of Goi\'{a}s, 74690-900, Goi\^{a}nia, GO, Brazil}
\affiliation{Department of Physical Chemistry, University of the Basque Country UPV/EHU, Apartado 644, E-48080 Bilbao, Spain}
\author{P. H. Souto Ribeiro}
\email{p.h.s.ribeiro@ufsc.br}
\affiliation{Departamento de F\'{i}sica, Universidade Federal de Santa Catarina, CEP 88040-900, Florian\'{o}plis, SC, Brazil}

\begin{abstract}
Photon pairs produced in spontaneous parametric down-conversion are naturally entangled in their transverse spatial
degrees of freedom including the orbital angular momentum. Pumping a non-linear crystal with a zero order Gaussian mode  
produces quantum correlated signal and idler photons with equal orbital angular momentum and opposite signs. Measurements 
performed on one of the photons prepares the state of the other remotely. We study the remote state preparation in this system 
from the perspective of  its potential application to Quantum Thermodynamics. 
\end{abstract}

\maketitle

\section{Introduction}

Light beams can have orbital angular momentum (OAM)  \cite{Yao:11,Padgett2011} and two or more photons can be prepared in entangled states of this degree of freedom  \cite{mair01}. OAM is a discrete and infinite dimension degree of freedom, making it suited for applications in classical and quantum communications \cite{Gibson:04}, among others. At the same time, Quantum Thermodynamics \cite{Anders_2017} is a growing field that explores the connections between Thermodynamics and Classical and Quantum Information theories, where the experimental platforms are still rare. The use of all optical schemes \cite{Zanin19} is highly attractive due to the available technologies for manipulating and controlling several degrees of freedom of light. In particular, the use of OAM mode was recently used \cite{Araujo18,Ribeiro19} to demonstrate the two-measurement protocol for studying the quantum version of Jarzynski's fluctuation relation  \cite{jar1997,Tasaki}.

The role of quantum correlations and coherence in Quantum Thermodynamics has been approached in recent works \cite{Tacchino18,Micadei19,Santos19}. Despite these initiatives, it is still unclear the range of advances that quantum correlations, entanglement and coherence can bring to the understanding of the basic concepts and potential applications in this field. Here, we produce photon pairs in spontaneous parametric down-conversion that are naturally entangled in OAM and use them in the remote state preparation of single-photon thermal states. Thermal statistics is actually present in the vast majority of light sources, like the sunlight, firelight and incandescent light bulbs. However, this statistics concerns the number of photons in an optical mode \cite{Arecchi65}. What we demonstrate here is the preparation of heralded single photon states presenting thermal statistics concerning the population probability of OAM modes. Instead of an optical single mode with several photons, as it is usual in Quantum Optics, we deal with single photons populating several modes.  

The thermal states prepared remotely might become a useful tool for the study of Quantum Thermodynamics using single photons as the working substance. We illustrate this potential utility analyzing a few processes applied to these thermal states. Another possibility using this scheme is the remote preparation of a thermal-like state possessing coherences. These are states with the same populations as a Gibbs state, but having also coherences. We discuss possible applications of this kind of state.

\section{State preparation}
Spontaneous parametric down-conversion (SPDC) is a nonlinear optical process in which a pump photon is converted into two photons usually called signal and idler. The quantum state describing signal and idler is given by a pure state \cite{mandel95} and  can be written in a simplified form, as an entangled state of the OAM degrees of freedom \cite{spiral}:

\begin{equation}
\vert \psi \rangle = \sum_{\ell_{a}, \ell_{b}} C^{\ell_{a}, \ell_{b}}_{\ell_p} \,\, \vert \ell_{a}, \ell_{b} \rangle,
\label{eq1}
\end{equation}
where $\ell_{p}$, $\ell_{a}$ and $\ell_{b}$ are the pump, signal and idler OAM indices, respectively. $C^{\ell_{a}, \ell_{b}}_{\ell_p}$ are coefficients accounting for the phase matching function and the OAM of the pump beam. 

We are interested in the particular case where the radial order is always $p=0$, and $\ell_p = 0$, so that the state in Eq. (\ref{eq1}) can be well approximated by 

\begin{equation}
\label{SPDC1}
\vert \psi \rangle = \sum_{\ell = -\infty}^{\infty} C_{|\ell|}  \,\, \vert +\ell, -\ell \rangle_{ab},
\end{equation}
where $C_{|\ell|} \equiv C^{\ell_{a}=+\ell, \ell_{b}=-\ell}_{\ell_p = 0}$. This is an infinite dimensional entangled state where signal and idler modes have the same $\ell$ with opposite signs. It is known that the coefficients $C_{|\ell|}$ decrease with increasing $\vert \ell \vert$ following an exponential decay law \cite{spiral}. Our aim is to identify the state of Eq. (\ref{SPDC1}) with a {\em nonlocal thermal} state of the form 
\begin{equation}
\vert \psi \rangle_{ab} = \sum_{\ell = -\infty}^{\infty} \sqrt{\frac{\mbox{e}^{-\beta \epsilon_{\ell}}}{z} } \,\, \vert +\ell, -\ell \rangle_{ab},
\label{eq:SPDCOAML0}
\end{equation}
where $\epsilon_{\ell} = (\vert\ell\vert + 1)\hbar\omega$ is the energy and the coefficient $\mbox{e}^{-\beta (\vert\ell\vert +1)\hbar\omega/2}/\sqrt{z}$ takes into account the normalization of the quantum state by means of the ``partition function'' $z$. The factor $(\vert\ell\vert + 1)\hbar\omega$ gives the energy as long as the radial number vanishes ($p=0$). Therefore, $\beta(\vert\ell\vert + 1)\hbar\omega$ is the ratio between the quantum of orbital angular momentum (in units of energy) and the thermal energy $k_B T = \beta ^{-1}$ (see the discussion presented in Ref. \cite{Araujo18}, where this identification is justified in detail). The global square root is introduced because this is an amplitude and not a probability. Eq. (\ref{eq:SPDCOAML0}) is a sort of thermally weighted entangled state. By tracing over idler (signal) photon OAM degrees of freedom, one obtains a \emph{remotely prepared local thermal state} in the signal (idler) beam, in the form
\begin{equation}
\rho^{th}_{a(b)} = \sum_{\ell = 0}^{\infty} \frac{\mbox{e}^{-\beta (\vert\ell\vert +1)\hbar\omega}}{z}  \,\, \vert \ell \rangle\langle \ell \vert_{b(a)}.
\label{eq:Thermal}
\end{equation}

Eq. (\ref{eq:Thermal}) shows that the spiral bandwidth defined in Ref. \cite{spiral} allows the interpretation of the bipartite SPDC entangled state as a source of local thermal states given that the OAM of one of the parts is traced out. In the following we present an experiment demonstrating this remote state preparation.

\section{Experiment}

\begin{figure}[h]
\includegraphics[width=0.48\textwidth]{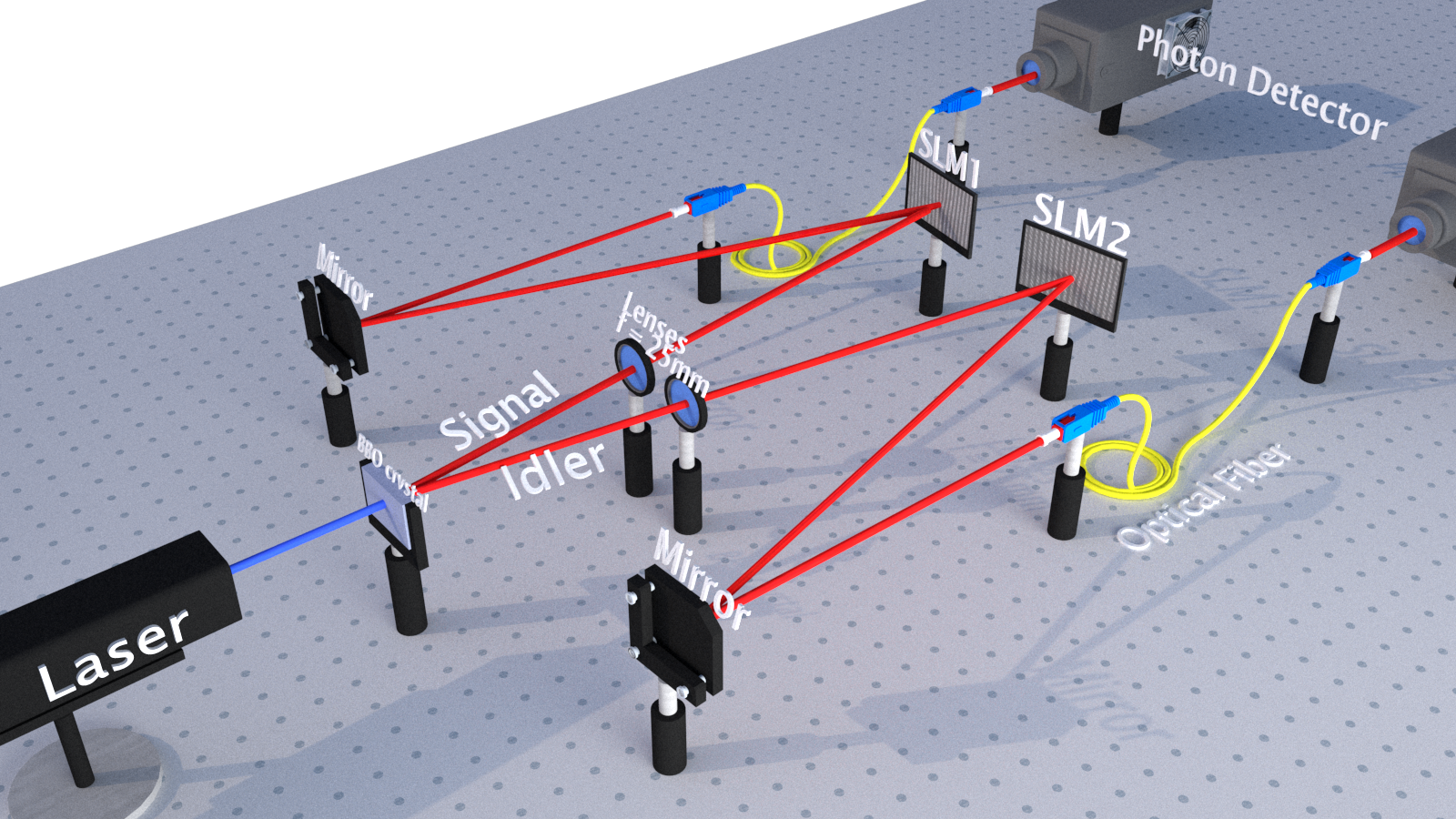}
\caption{Sketch of the experimental set-up. After down-conversion, signal and idler photons are directed to SLMs and afterwards coupled to single-mode optical fibers, and then detected by single-photon counting modules (SPCM). Coincidences are registered between signal and idler detection events. }
\label{fig:setup}
\end{figure}

Figure \ref{fig:setup} shows the sketch of the experimental set-up for demonstrating the remote preparation of {\em single photon vortex thermal states}.  A 405~nm diode-laser pumps a Beta Barium Borate (BBO) crystal and spontaneous parametric down-conversion takes place according to type I phase matching. Signal and idler beams propagate through collimating lenses with focal length $f$ = 25~mm situated at 25~mm from the crystal. 
Signal and idler beams are directed to spatial light modulators (SLM), where their wavefronts can be modulated. When the applied modulation is such that the OAM of the input beam is raised or lowered, the output photon can eventually be coupled to a zero order ($\ell = 0$) OAM optical mode. In this case, and only in this case, the photon will be coupled to a single-mode optical fiber. The output of the fibers in the signal and idler modes are then coupled to a single-photon-counting-module (SPCM). Notice that this scheme allows the measurement of the OAM of signal and idler photons by scanning the SLMs, rising and lowering the OAM and counting the number of photons coupled to the fiber for each applied mask.
For instance, applying a mask ${\cal L}_{+m}$ that performes the raising operation $\ell  \rightarrow  \ell+m$ detects photons in modes with $\ell = -m$. 

For the generation of thermal states, say in the signal beam, we should trace over the idler OAM degres of freedom. In this case, the idler SLM is disabled and the optical fiber
is replaced with a bucket detector. The OAM measurement scheme is used only for the signal photons. When an idler photon count happens, the presence of a signal photon is heralded and prepared in a thermal state like Eq. (\ref{eq:Thermal}).

The scheme with OAM measurements in both signal and idler photons can be used for preparing another type of heralded single photon state, given by:

\begin{equation}
\vert \psi \rangle = \sum_{\ell = -\infty}^{\infty} \frac{\mbox{e}^{-\beta \epsilon_{\ell}}}{z}  \,\, \vert \ell \rangle,
\label{eq:ch_therm}
\end{equation}
where $\epsilon_{\ell} = (\vert\ell\vert +1)\hbar\omega$.

The preparation of this state is achieved by using the idler SLM to apply superpositions of raising and lowering operations with appropriate
weights:

\begin{equation}
{\cal L} \rightarrow \sum_{m} c_m ({\cal L}_{+m} + {\cal L}_{-m}) ,
\label{eq2:ch_therm}
\end{equation}
where the coefficients $c_m = \frac{\mbox{e}^{-\beta \epsilon_{m}}}{z}$ follow a thermal distribution. 
This generates a remotely prepared signal photon, which is a pure state having the same populations
(i.e. the diagonal elements of the density matrix in the energy basis) as a thermal state but also having coherences. 
We suggest that this kind of state can be useful as a resource in Quantum Thermodynamics processes, in the same fashion as negative temperature \cite{Assis19}
and squeezed \cite{Klaers17} heat baths.

\section{Results}

\subsection{Remote preparation of single photon vortex thermal states} \label{prep}

\begin{figure}
\centering
\includegraphics[width=0.45\textwidth]{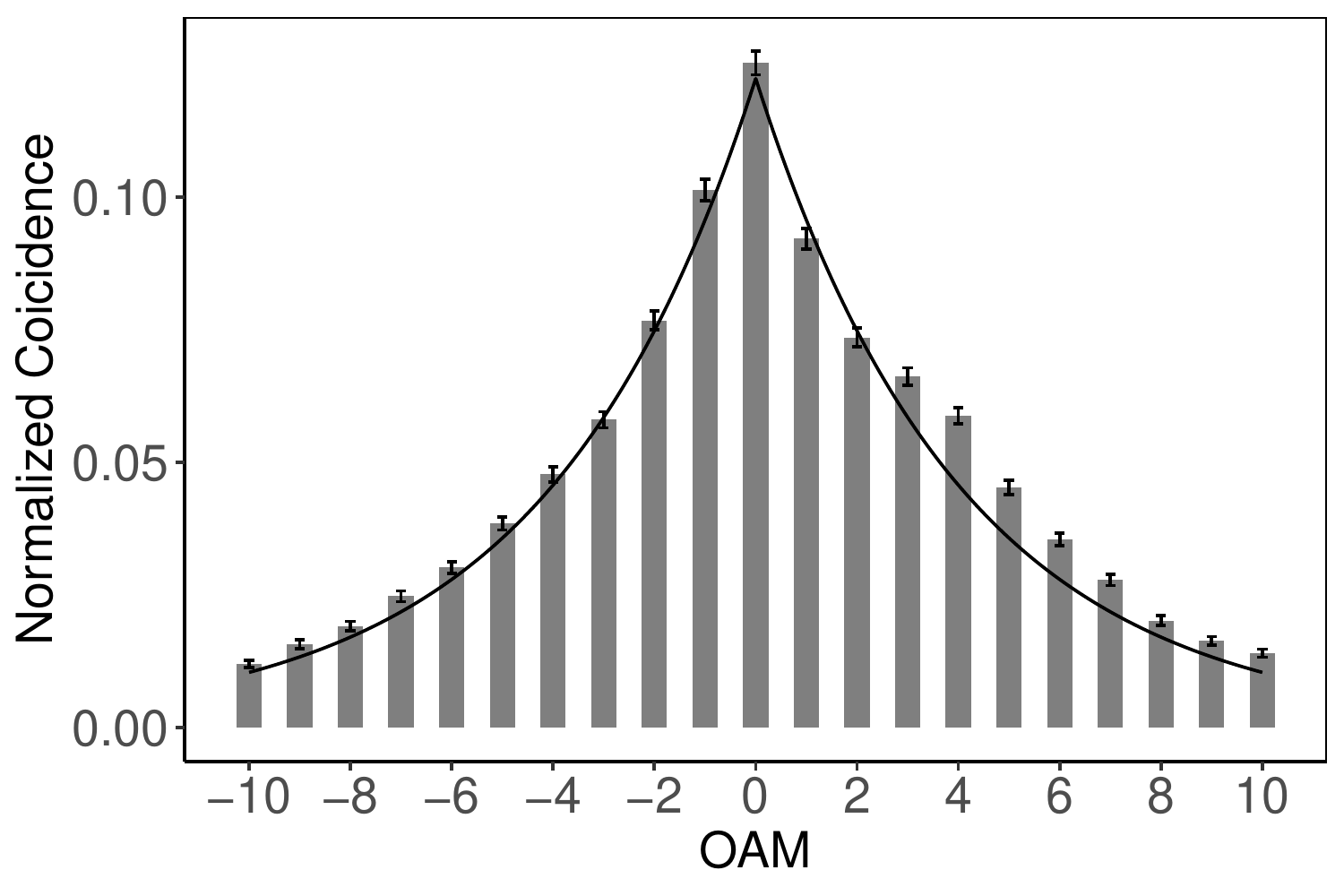}
\caption{OAM distribution for heralded idler photons gated by signal photons with traced OAM. A thermal distribution was fitted to the experimental data (solid line).}
\label{fig:measurement_dist1}
\end{figure}

Figure \ref{fig:measurement_dist1} shows a normalized distribution of coincidence counts between signal and idler photons as a function of the OAM. This results were obtained by pumping the crystal with an ordinary Gaussian laser mode that has $\ell = 0$ and performing the detection of the idler photons with a bucket detector, which traces out the OAM. The signal photons were directed to the SLM and then to the fiber, so that the OAM measurement scheme was applied. Therefore, Figure \ref{fig:measurement_dist1} shows the OAM spectrum for the heralded signal photon.
As the photon counting statistics is Poissonian, the error bars are the standard deviation and therefore the square root of the coincidence counting rate.

Fitting the data to a thermal distribution like
\begin{equation}
    p(\ell) = \frac{e^{-\alpha (|\ell|+1)}}{Z} ,
    \label{eq:fit}
\end{equation}
where $\alpha = \beta\hbar\omega$. Defining $k_B = 1$, we can infer the effective temperature of the single photon state and obtain the with respect to $\hbar \omega$ normalized inverse thermal energy. 
This is not a Thermodynamic temperature. It concerns the parameter of the Gibbs state, which is usual for
describing thermal states of quantum systems.

Due to the degeneracy for all energy levels except $\ell = 0$, we can write the partition function as $Z^{-1} = \left(e^{\alpha} \tanh{\frac{\alpha}{2}} \right)$. 
In Fig. \ref{fig:measurement_dist1} this fit to the thermal function is displayed by the solid line. This result shows that the measurements are compatible with a thermal distribution, and that the positive and negative parts are approximately symmetric. This distribution was found to have $\alpha \approx 0.25$. 

\begin{figure}
\centering
\includegraphics[width=0.45\textwidth]{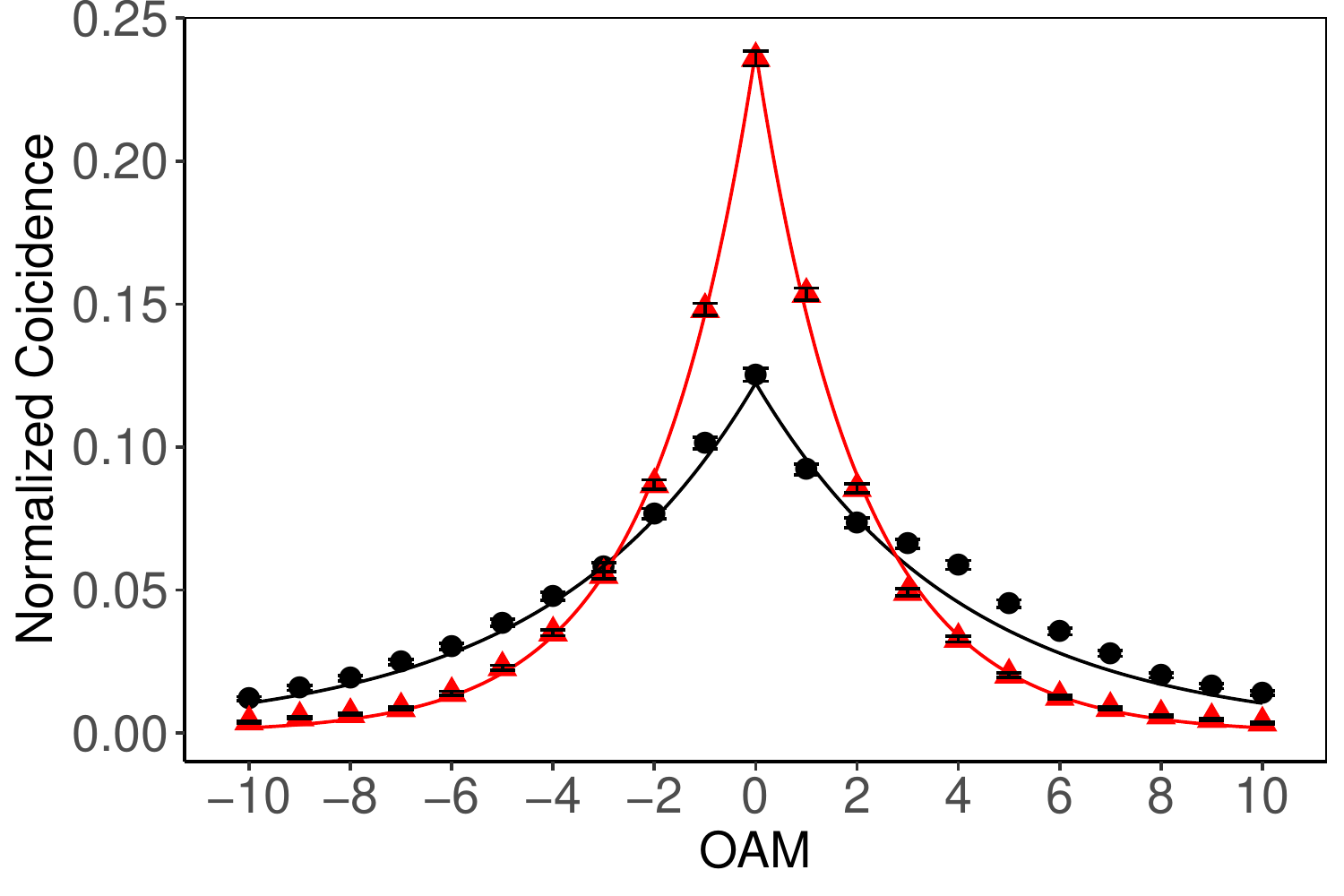}
\caption{OAM distribution for heralded idler photons without (black circles) and with reduced pump beam diameter (red triangles). Thermal functions were fitted to the experimental data (solid lines).}
\label{fig:measurement_dist2}
\end{figure}

According to Ref. \cite{spiral}, the decay rate of the exponential can be changed by varying the diameter of the pump
beam inside the crystal. This is equivalent of tuning the effective temperature.
In order to demonstrate it, we have used an optical telescope in the pump beam before the crystal in order to decrease its diameter. 
This change is supposed to decrease the effective temperature of the thermal distribution.
In Fig. \ref{fig:measurement_dist2} we show a comparison between the OAM distributions for the two beam diameters/temperatures. 

The ratio between the diameters is 3  and  the effective temperature changed from $\alpha_1 \approx 0.25$ to $\alpha_2 \approx 0.49$, recalling that $\alpha$ is a parameter linked to the inverse temperature, so that decreasing the beam diameter increases $\alpha$ and cools down the photonic mode structure. We also compute the average dimensionless energy defined as

\begin{equation}
    \frac{\langle E \rangle}{\hbar\omega} = \frac{1}{\hbar\omega} \sum_{\ell} p(\ell) \varepsilon (\ell)~,
    \label{eq:energy}
\end{equation}
where $p(\ell)$ is the measured probability for each $\ell$ and $\varepsilon (\ell)= (\vert{\ell} \vert +1)\hbar \omega$ is the energy for each $\ell$.

The energies for the two different beam diameters are  $\langle E_1 \rangle /\hbar\omega \approx$ 4.34 and $\langle E_2 \rangle /\hbar\omega \approx$ 3.00.

\section{Effect of turbulence}\label{sec:turbo}

\begin{figure}
\centering
\includegraphics[width=0.45\textwidth]{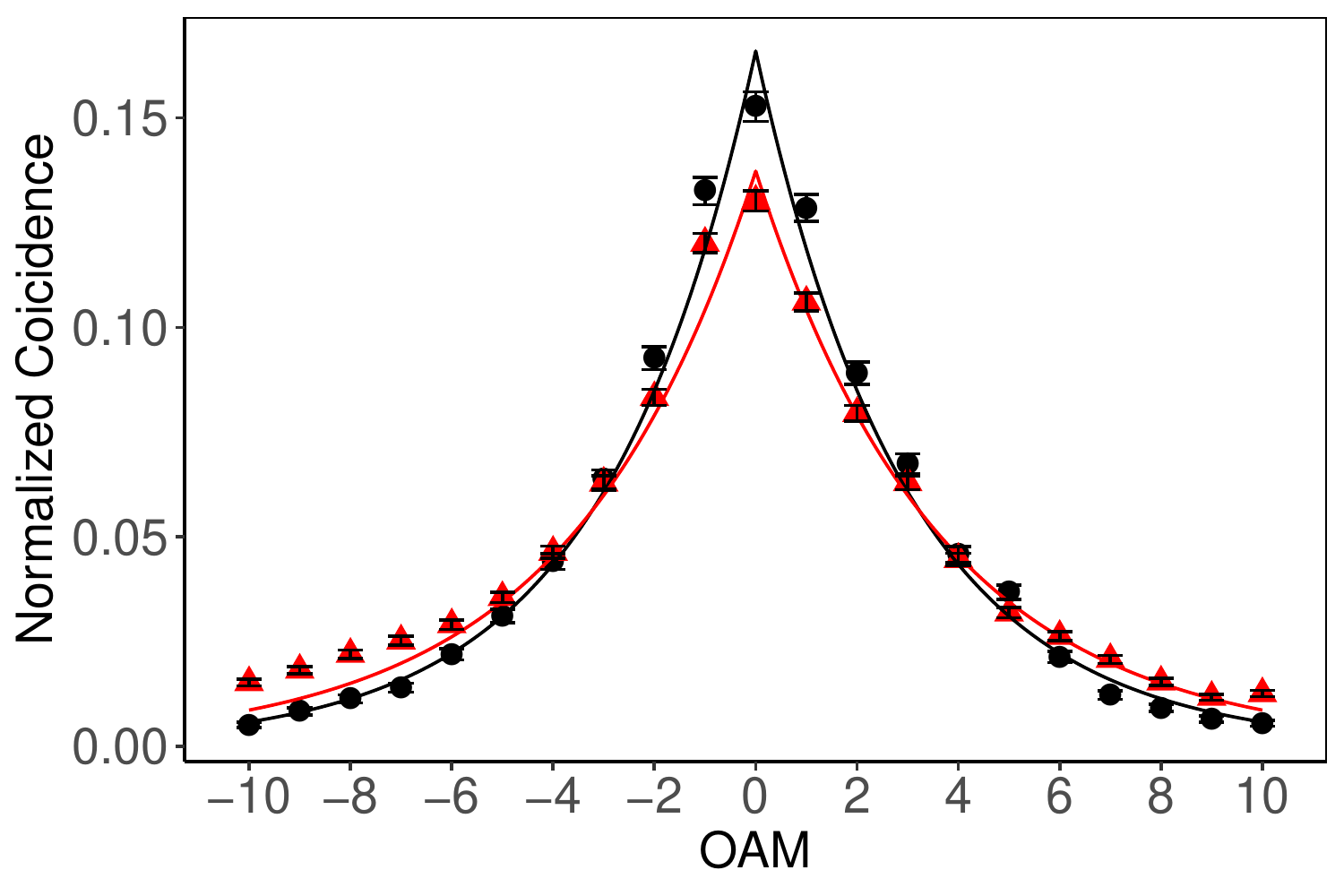}
\caption{Measured OAM distribution for heralded idler photons without (black circles) and with (red triangles) a weak atmospheric turbulence. A thermal distribution was fitted to the experimental data.}
\label{fig:turbo1}
\end{figure}

We would like to illustrate the utility of a single photon thermal state by submitting it to some physical process. Being an optical system, the propagation through a turbulent atmosphere
that couples different input and output OAM modes is an interesting process. The propagation through a turbulent atmosphere can be simulated with a SLM \cite{Rodenburg14}
using a phase mask designed to reproduce Kolmogorov turbulence. Figure \ref{fig:turbo1} shows the recorded distribution with and without a weak turbulence process. The input thermal distribution has an inverse temperature $\alpha_1 \approx 0.34$. The distribution after the process is slightly out of equilibrium and in order to provide some figure of merit concerning the deviation with respect to the equilibrium state, we calculate the {\em Kullback-Leibler (KL) divergence}, which is defined as

\begin{equation}
D_{KL} \left( p_f (\ell) || p_m (\ell) \right) = - \sum_{\ell} p_m (\ell) \log \left( \frac{p_m (\ell)}{p_f (\ell)} \right)~,
\label{TD}
\end{equation}
where ${ p_m (\ell)}$  and ${p_f (\ell)}$ are probability distributions. We calculate this divergence between our recorded distribution and the probabilities taken from the thermal state fitted to that distribution. We can therefore quantify the divergence between our measured distributions and the respective thermal equilibrium state.

The distributions are truncated at $\ell$ = 10. For the input state, KL divergence is $D_{KL}(i) \approx 0.015$ meaning that it is approximately a thermal state, while for the output distribution $D_{KL}(o) \approx 0.027$. This shows that the output state is not very far from a thermal state in what concerns the populations. In fact, the process may have created superpositions of states, which would take the system out from equilibrium. The resulting OAM distribution is fitted to a thermal distribution and an inverse temperature $\alpha \approx$ 0.28 is found. Therefore, the turbulence seems to heat up the system, as intuitively expected. In order to see this, we recall that a photon at zero temperature, or infinite $\beta$, corresponds to a pure and very clean Gaussian mode, while a hot photon corresponds to a distribution with considerable contributions from several
higher order modes.

We have also computed the average energies.
For the input state, we find  $ \langle E \rangle \approx 3.56 \hbar\omega$ and after the process $ \langle E \rangle \approx 4.13 \hbar\omega$. We can see that the process produces a small deviation from the equilibrium state and a small change in the average energy.

\begin{figure}[t!]
\centering
\includegraphics[width=0.48\textwidth]{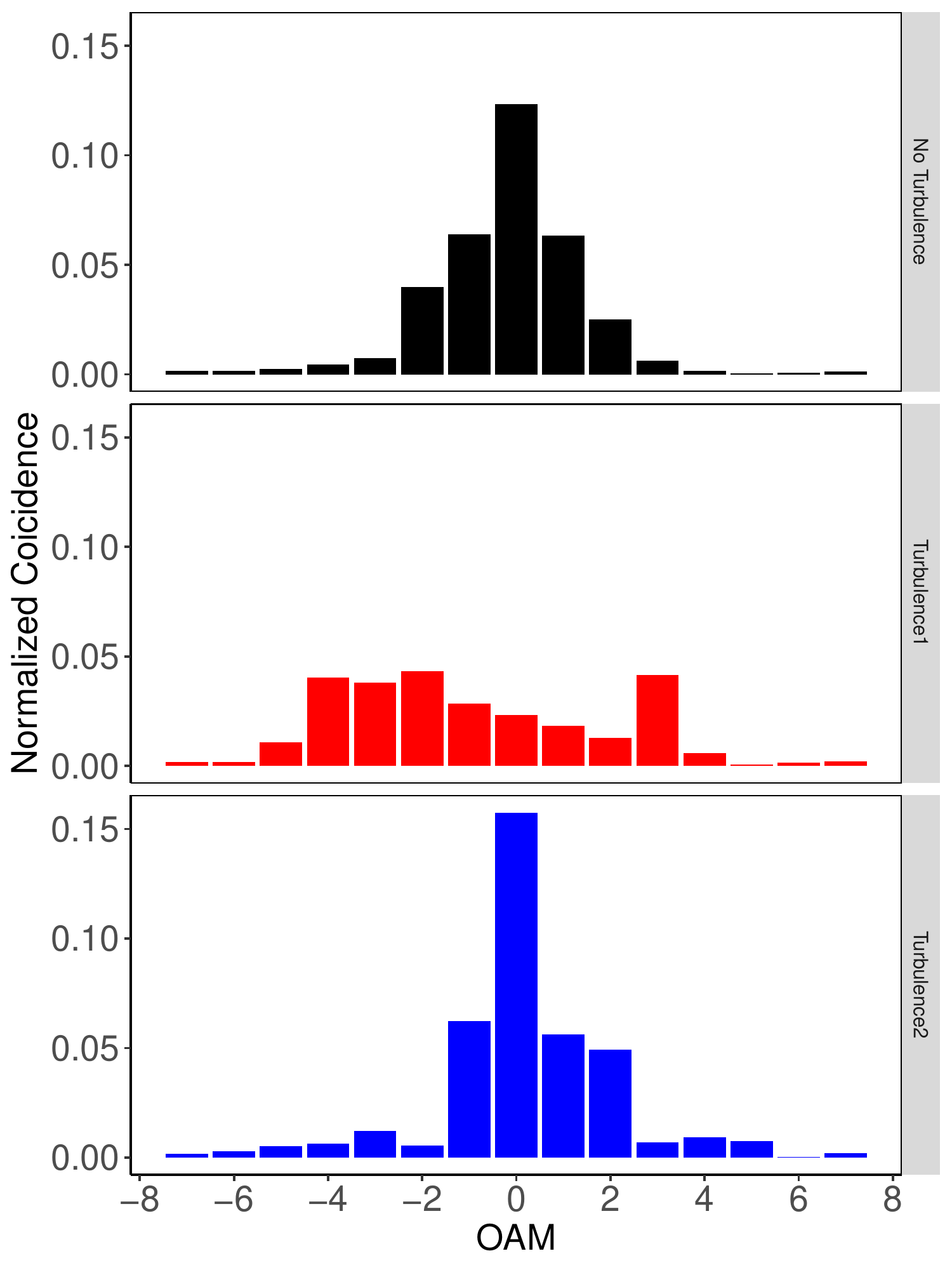}
\caption{Measured OAM distributions for the prepared \textit{coherent} thermal state, without (black) and with two different turbulences of the same strength (red and blue).}
\label{fig:turbo2}
\end{figure}

We apply the same process to the \textit{coherent} thermal state (Eq. (\ref{eq:ch_therm})) when the idler SLM applies a mask realizing the operation described in Eq. (\ref{eq2:ch_therm}). The process of turbulence produces a much stronger effect on this state as compared to the thermal one. This is probably due to interference effects 
allowed by the coherences in the state. Figure \ref{fig:turbo2} shows the unperturbed input \textit{coherent} thermal distribution and two measured distributions for different turbulence masks but the same scintillation strengths. The scintillation strength is related to a parameter in Komolgorov turbulence that essentially controls the length scale of the phase modulations in the turbulence mask. It is possible to generate several different masks with the same scintillation strength.

\begin{figure}[t!]
\centering
\includegraphics[width=0.48\textwidth]{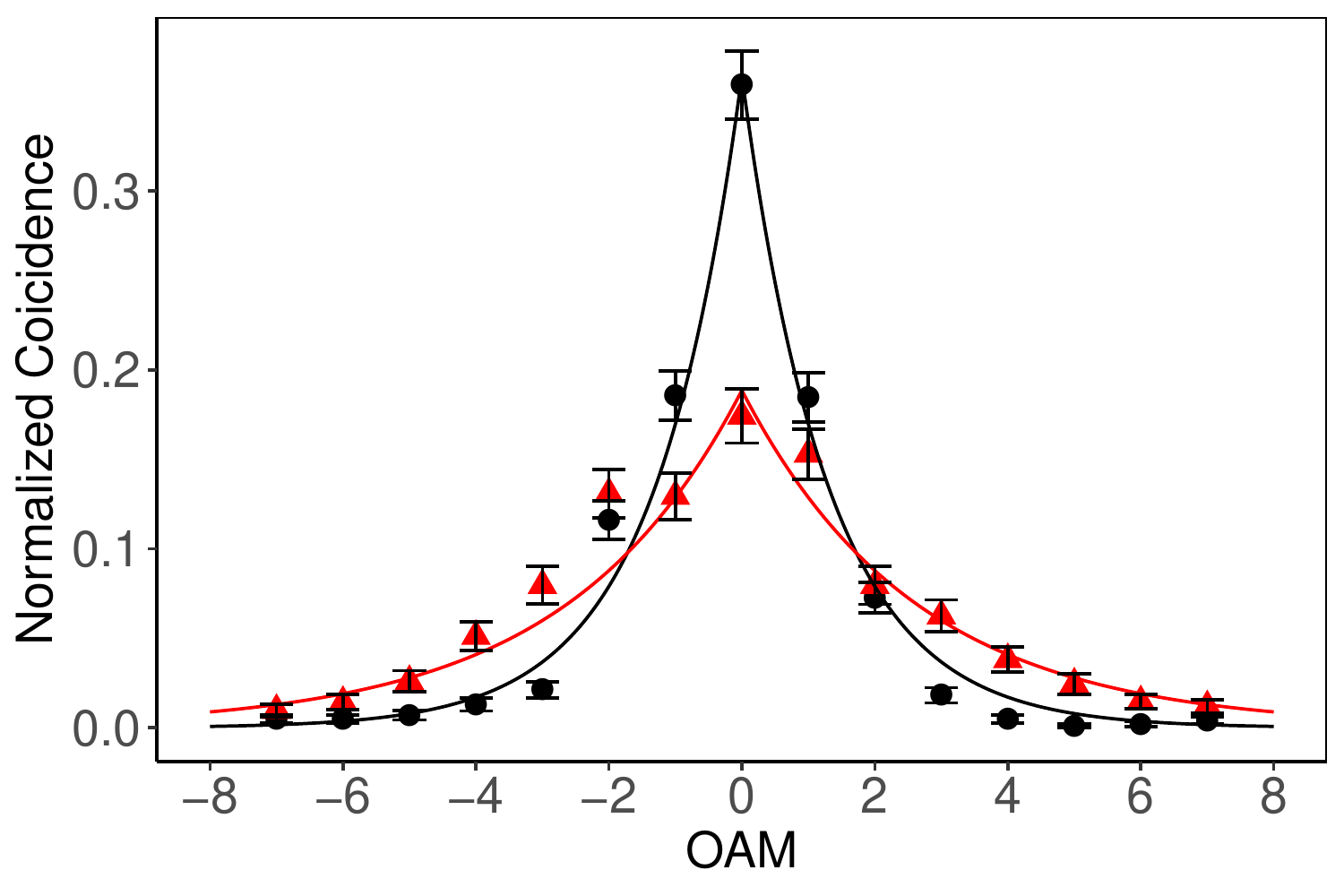}
\caption{OAM distribution for the heralded signal photon of the unperturbed {\em coherent} thermal state (discs). OAM distribution after averaging over several measurements with different turbulence masks of the same scintillation strength (triangles).}
\label{fig:turbo3}
\end{figure}

Applying turbulence masks to the coherent thermal state using several different turbulence masks of the same strength and summing over the recorded distributions, we note that the state tends to return to a thermal state, as we can see in Fig. \ref{fig:turbo3}. These results were obtained after summing the measured distributions for 10 different turbulence masks. Fitting a thermal distribution to the recorded data without and with applied turbulence results in inverse temperatures of $\alpha_1 \approx 0.76$ and $\alpha_2 \approx 0.38$ respectively.

\section{Measurements with different sizes of the idler detector aperture}\label{sec:aperture}
  
\begin{figure}[!t]
\centering
\includegraphics[width=0.48\textwidth]{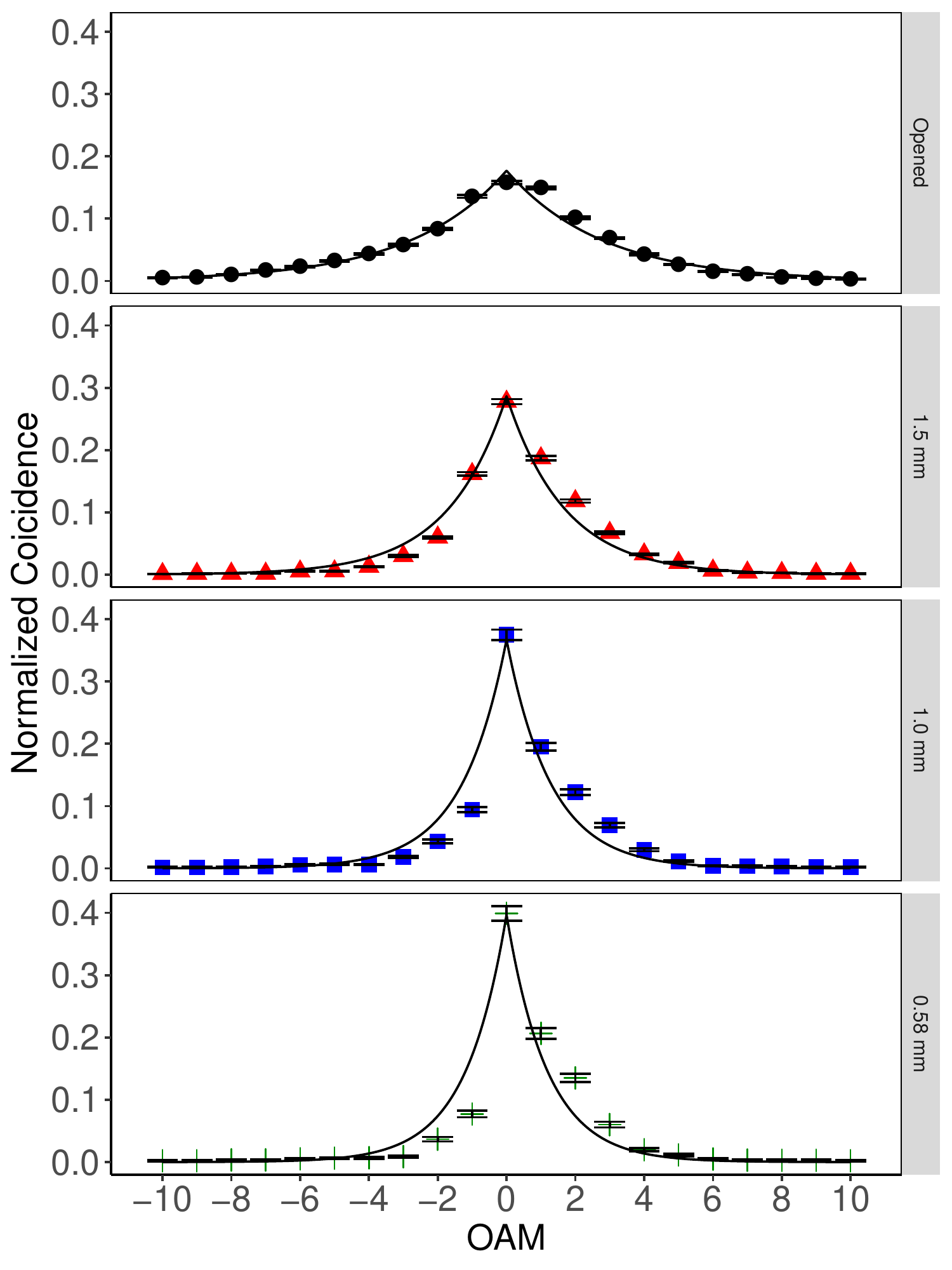}
\caption{OAM distributions for different aperture sizes before the photon detector; opened detector (discs), 1.5~mm (triangles), 1.0~mm (squares), 0.58~mm (crosses). Thermal distributions were fitted to the experimental data (solid lines). }\label{fig:aperture}
\end{figure}

In the measurements presented so far the idler detector was completely opened and therefore, the trace over the idler OAM states is performed. However, if the detector is partially opened, the tracing operation is also partial and the modes with higher OAM are not detected. We have analyzed the effect of the partial tracing by placing an iris in front of the idler detector and varied the detector's diameter. We have used diameters of $1.5$ mm, $1.0$ mm and $0.58$ mm. 
Figure \ref{fig:aperture} shows the results for these measurements.

The results show that all distributions still follow the shape of a thermal distribution and that the smaller the diameter, the lower the temperature. In the limit case, where the idler detector is practically closed, the system temperature will tend to zero (pure state with the only contribution $\ell = 0$).
This behavior is explained by the fact that there is a direct correlation between signal and idler OAM and the collection of lower OAM in the idler should result in a colder distribution in the signal.

The KL divergence, average energies and inverse temperatures for the four distributions in Figure \ref{fig:aperture} are
a) detector completely opened (discs): $t_d \simeq 0.014$,  $ \langle E \rangle \simeq 3.40\hbar\omega$, $\alpha \simeq 0.35$; b) 1.5 mm aperture (triangles): $t_d \simeq 0.015$, $ \langle E \rangle=2.51 \hbar\omega$,
$\alpha \simeq 0.59$; c) 1.0 mm aperture (squares): $t_d \simeq 0.036$, $ \langle E \rangle=2.33 \hbar\omega$, $\alpha \simeq 0.77$; 
d) 0.58 mm aperture (crosses): $t_d \simeq 0.054$, $ \langle E \rangle=2.23 \hbar\omega$, $\alpha \simeq 0.84$. 

This analysis indicates that there is another convenient control parameter for the temperature of the heralded signal photon OAM distribution besides the diameter of the pump beam in the
crystal.

\section{Conclusion}
In conclusion, we present a method for remote state preparation of  single photon (vortex) thermal states based on the orbital angular momentum degree of freedom. 
We demonstrate experimentally that the heralded single photon state has an exponentially decaying structure that can be adjusted to a thermal distribution. We also show that the temperature can be tuned by changing the diameter of the pump beam inside the non-linear down-conversion crystal or the diameter of the idler detector in the non-local process.
We submit the prepared state to a turbulence processes and analyze the output state. The turbulence performs work on the system and drives it out of equilibrium.

This study advances the knowledge on how to use this optical system to experimentally study Quantum Thermodynamics. 

\begin{acknowledgements}
The authors would like to thank the Brazilian Agencies CNPq, FAPESC, FAPEG, and the Brazilian National Institute of Science and Technology of Quantum Information (INCT/IQ). This study was financed in part by the Coordena\c{c}\~ao de Aperfei\c{c}oamento de Pessoal de N\'{i}vel Superior - Brasil (CAPES) - Finance Code 001. LCC would like to also acknowledge support from Spanish MCIU/AEI/FEDER (PGC2018-095113-B-I00), Basque Government IT986-16, the projects QMiCS (820505) and OpenSuperQ (820363) of the EU Flagship on Quantum Technologies and the EU FET Open Grant Quromorphic and the U.S. Department of Energy, Office of Science, Office of Advanced Scientific Computing Research (ASCR) quantum algorithm teams program, under field work proposal number ERKJ333.
\end{acknowledgements}

\end{document}